\documentclass[twocolumn,showpacs,preprintnumbers,amsmath,amssymb]{revtex4-1}

\usepackage{graphicx}
\usepackage{dcolumn}
\usepackage{bm}
\usepackage{color}

\begin{document}

\title{Competing Turing and Faraday instabilities in longitudinally modulated\\ passive resonators}

\author{Fran\c cois Copie}
\email{francois.copie@univ-lille1.fr}
\author{Matteo Conforti}
\author{Alexandre Kudlinski}
\author{Arnaud Mussot}
\affiliation{
Univ. Lille, CNRS, UMR 8523 - PhLAM - Physique des Lasers Atomes et Mol{\'e}cules, F-59000 Lille, France
}

\author{Stefano Trillo}
\affiliation{
Department of Engineering, University of Ferrara, Via Saragat 1, 44122 Ferrara, Italy
}

\begin{abstract}
We experimentally investigate the interplay of Turing (modulational) and Faraday (parametric) instabilities in a bistable passive nonlinear resonator. The Faraday branch is induced via parametric resonance owing to a periodic modulation of the resonator dispersion. We show that the bistable switching dynamics is dramatically affected by the competition between the two instability mechanisms, which dictates two completely novel scenarios. At low detunings from resonance switching occurs between the stable stationary lower branch and the Faraday-unstable upper branch, whereas at high detunings we observe the crossover between the Turing and Faraday periodic structures.  
The results are well explained in terms of the universal Lugiato-Lefever model.

\end{abstract}


\maketitle

\textit{Introduction}.-- Dissipative spatial (localized and patterned) structures  are ubiquitous in nonlinear extended biological, chemical and physical systems operating far from equilibrium \cite{cross_pattern_1993, cross_pattern_2009, descalzi_localized_2011, Tlidi_Localized_2014}. A universal triggering mechanism is the Turing (modulational) instability, which, owing to a symmetry breaking bifurcation, favors the growth of modulations against the homogeneous solutions \cite{castets_experimental_1990, lugiato_spatial_1987}. This growth eventually saturates, giving rise to patterns with intrinsic wavenumbers which attract the dynamics.
On a completely different basis, in systems forced at frequency $\omega_d$, the characteristic wavenumber $k$ is selected via the dispersion relationship and a 2:1 parametric resonance to fulfill the relation $\omega(k)=\omega_d/2$ (generally also multiple $2m:1$ resonances, $m$ integer, are possible in unstable regions known as Arnold tongues \cite{lin_resonant_2000}).
This phenomenon, discovered by Faraday in a vertically vibrating fluid \cite{faraday_peculiar_1831} and explained much later \cite{benjamin_stability_1954, coullet_dispersion-induced_1994}, is at the origin of Faraday waves observable in different areas of physics \cite{melo_transition_1994, szwaj_faraday_1998, longhi_nonadiabatic_2000, staliunas_faraday_2002, engels_observation_2007}. While the Turing and Faraday mechanisms have different physical origins, they can in principle coexist. However, the observation of the effects of their competition is surprisingly lacking.\\
\indent In this letter, we consider a bistable system and report experimental evidence for the fact that such competition drastically changes its switching dynamics. We employ a passive fiber resonator where such instabilities manifest themselves in time domain \cite{haelterman_additive-modulation-instability_1992, haelterman_dissipative_1992, coen_modulational_1997, coen_experimental_1998}. Passive microresonators and fiber rings, all described by transpositions of the universal Lugiato-Lefever equation (LLE) introduced in the spatial case \cite{lugiato_spatial_1987},  have indeed proven extremely effective for the observation of temporal dissipative structures such as solitons and primary frequency combs \cite{delhaye_optical_2007, kozyreff_localized_2009, leo_temporal_2010, leo_nonlinear_2013, herr_temporal_2014, jang_temporal_2015, pfeifle_optimally_2015, xue_mode-locked_2015}. In our experiments, we engineer the resonator to have periodic group velocity dispersion (GVD) with {\em normal} average value. The GVD management acts as a spatial forcing, while the normal GVD guarantees high gain for the Faraday branch \cite{conforti_modulational_2014}, even though it makes the excitation of Turing structures much more critical \cite{coen_experimental_1998, hansson_dynamics_2013}.
This regime allows us to give spectral evidence for the spontaneous formation of periodic structures, which follows two different novel scenarios: (i) at relatively small detunings from resonance, switching occurs from homogeneous state to Faraday structures, with the Turing instability only acting as a trigger; (ii) crossover from periodic structures of the Turing and Faraday types can occur at large detunings by controlling the pump power. Such results pave the way towards the control and taming of the instabilities via induced periodicity in a variety of different settings \cite{staliunas_parametric_2013, conforti_modulational_2014, kumar_taming_2015}.

\begin{figure}
\includegraphics[width=.47\textwidth]{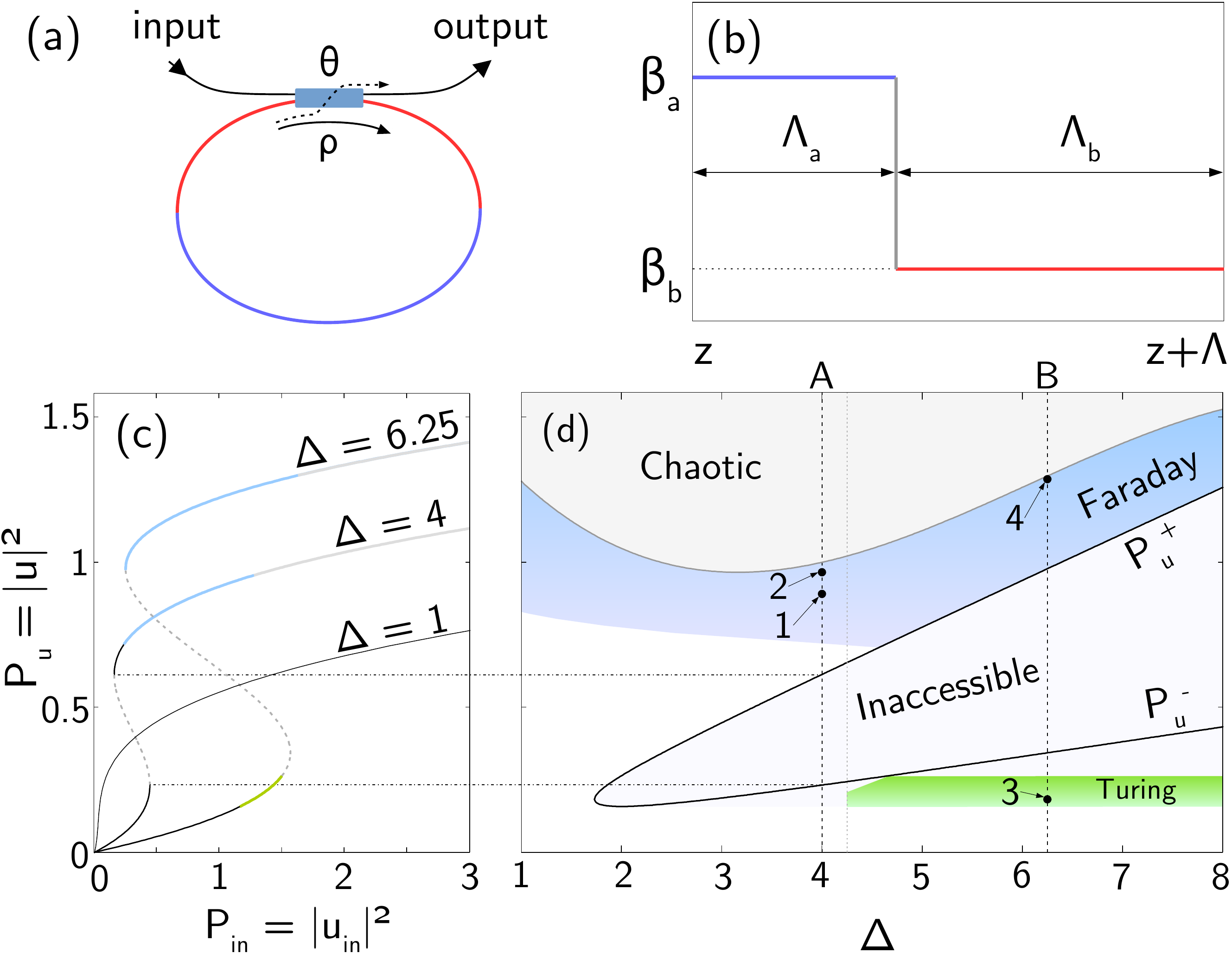}
\caption{\label{fig:mapping} (a) Schematic illustration of a non-uniform passive fiber ring cavity. (b) Piecewise constant dispersion map over one period of the GVD. (c) Normalized steady-state curves for detuning values $\Delta = 1, 4$ and $6.25$. (d) Instability domains in a cavity with dispersion map as in (b) in the plane ($\Delta$,$P_u$). The ``Inaccesible'' region corresponds to the negative-slope branch of the steady-state. The green and blue domains represent the regions where Turing and Faraday modulated structures can be excited (at higher powers the Faraday structures destabilize giving rise to chaotic spatio-temporal evolutions). The bullets labelled 1,2,3 and 4  indicate the corresponding experimental results of Fig. \ref{fig:exp1} and \ref{fig:exp2}. Parameters: $\Lambda_{a} = 0.97$, $\Lambda_{b} = 0.03$, $\beta_{a} = 1.5$, $\beta_{b} = -14$, $\alpha = 0.157$, $\theta^2 = 0.1$, $\Lambda = 1$: the period of the GVD equals the length of the resonator} 
\end{figure}

\textit{General behaviour}.-- We consider the passive fiber ring cavity sketched in Fig. \ref{fig:mapping}(a) with varying dispersion [Fig. \ref{fig:mapping}(b)], which is well modeled by the LLE \cite{lugiato_spatial_1987, conforti_modulational_2014}: 
\begin{equation}\label{eq:LLE}
i \frac{\partial u}{\partial z} - \frac{\beta(z)}{2} \frac{\partial^2 u}{\partial t^2} + \lvert u \rvert ^2 u = (\delta - i \alpha) u + i S,
\end{equation}
where $z = Z/L$, $t = T/T_0$, $u(z, t) = E(Z, T) \sqrt{\gamma L}$, $\beta(z) = \beta_2(z) / \lvert \beta_{2}^{av} \rvert$. $L$ is the cavity length, $T_0 = \sqrt{\lvert \beta_{2}^{av} \rvert L}$, $\beta_2(z)= d^2 k / d \omega^2$ is the periodic GVD, with average value $\beta_{2}^{av}$, and $\gamma$ is the fiber nonlinearity. $Z$, $T$ and $E$ denote real-world distance, group velocity delayed time, and the intracavity field envelope respectively. $S$ is the driving term such as  $S = \sqrt{P} = \theta u_{in}$, where $\theta$ is the coupler transmission coefficient ($\rho^2 + \theta^2 = 1$) and $u_{in} = \sqrt{\gamma L} E_{in}$ is the normalized input field.  $\delta$ is the cavity detuning and $\alpha$ describes the total losses (output coupling, linear and splicing losses). In the following, we will refer to the normalized detuning defined as $\Delta = \delta / \alpha$. 

The steady-state homogeneous solutions (i.e., $\partial_z=\partial_t=0$) of Eq. (\ref{eq:LLE}) can become unstable against the growth of optical modulations at frequency $\omega$, following either a Turing mechanism \cite{haelterman_dissipative_1992, haelterman_additive-modulation-instability_1992}, or a Faraday mechanism when forcing is present \cite{conforti_modulational_2014}. The most unstable frequency and the relative gain in the Turing case are \cite{haelterman_dissipative_1992, haelterman_additive-modulation-instability_1992,sup_mat}
\begin{equation}\label{eq:Turing}
\omega_{T}=\sqrt{\frac{2}{\beta_{av}}(\delta-2 P_u)},\;\; g(\omega_{T})= P_u -\alpha.
\end{equation}
where $\beta_{av}=\beta_2^{av}/|\beta_2^{av}|=1$ and $P_u=|u|^2$ are the normalized average GVD and intracavity power, respectively. Conversely, when forcing with period $\Lambda$ (periodicity of $\beta_2(z)$ in units of cavity length $L$) is present, it follows from Floquet theory that parametric (Faraday) instabilities set in around multiple frequencies \cite{conforti_modulational_2014,sup_mat}
\begin{equation}\label{eq:QPM}
\omega_{m} = \sqrt{\frac{2}{\beta_{av}} \left[ (\delta - 2P_u) \pm \sqrt{\left ( \frac{m \pi}{\Lambda} \right )^2 + P_u^2 } \right]},
\end{equation}
$m=1,2,....$ ($m=1$ in the experiments), which represent the tips of Arnold tongues. Importantly, at frequencies $\omega_{m}$ in Eq. (\ref{eq:QPM}), the perturbation wavenumber equals an integer multiple $m$ of half the forcing wavenumber $\pi/\Lambda$, which corresponds indeed to the parametric resonance condition \cite{conforti_modulational_2014,sup_mat} (with inverted role of $k$ and $\omega$ with respect to the spatial case mentioned before). The large difference of frequency between the sidebands of Turing ($\omega_{T}$) and Faraday ($\omega_{m}$) types is the distinctive trait which allows us to unequivocally identify the different regimes of instability in our experiments. Note that Turing instabilities are of different physical origin with respect to the ones in the normal GVD regime observed in conservative settings which do require the periodicity \cite{matera_sideband_1993, droques_experimental_2012, droques_dynamics_2013}. In order to understand how the controlling parameters, namely power and detuning, affect the behavior of the system, we summarize in Fig. \ref{fig:mapping}(d) the domains of the different instabilities in the parameter plane $(\Delta,P_u)$. For better clarity, we also report in Fig. \ref{fig:mapping}(c) the steady-state response for different values of detuning, namely $\Delta=1,4$ and $6.25$. For $\Delta \ge \sqrt{3}$, the cavity is bistable \cite{lugiato_spatial_1987, haelterman_additive-modulation-instability_1992}, and exhibits an unstable negative slope branch for $\text{P}_u^{-}<\text{P}_u<\text{P}_u^{+}$, where  $\text{P}_u^{\pm}(\Delta)$ stand for the 
bistability knees, delimiting the domain labelled ``Inaccessible'' in Fig. \ref{fig:mapping}(d). 

Below such domain, the green area corresponds to the region where temporally modulated Turing structures can be excited. This region has been computed numerically and corresponds to the tiny domain where Turing structures, which bifurcate subcritically, can be spontaneously formed \cite{coen_bistable_1999}. We emphasize that this regime requires to drive the cavity with a detuning $\Delta > 4.25$, and with powers belonging to a small portion of the lower branch of the bistable response (highlighted in green over the bistable curve for $\Delta = 6.25$ in Fig. \ref{fig:mapping}(c)). It is important to emphasize that this regime only depends on the average GVD and not on its periodic modulation and would thus also appear in uniform cavities. 
On the contrary, Faraday structures only develop when the cavity is driven over the upper branch and the periodic longitudinal variations are effective. As a result, the stable excitation of Faraday structures requires to operate in the blue domain of Fig. \ref{fig:mapping}(d). At higher powers, such structures destabilize leading to chaotic states (see upper portion of Fig. \ref{fig:mapping}(d)). Note that the Faraday branch (unlike the Turing one) extends also to the monostable regime $\Delta < \sqrt{3}$. However, in this letter, we focus on the bistable regime where the two instabilities can compete thereby drastically changing the bistable switching dynamics.
In particular, Fig. \ref{fig:mapping}(d) allows to envisage two distinct regimes that we experimentally address: (i) at low detuning ($\Delta < 4.25$, see vertical line A) switching occurs between the steady-state lower branch and an upper branch which is Faraday-unstable. In this case the Turing instability can only favour such switching, whereas no stable Turing structures can be created; (ii) at high detunings ($\Delta > 4.25$, see vertical line B),  switching between Turing and Faraday structures can be controlled by the power. A careful stabilization of the cavity allows us to accurately control the detuning and observe the scenarios (i) and (ii).

\begin{figure}
\includegraphics[width=.47\textwidth]{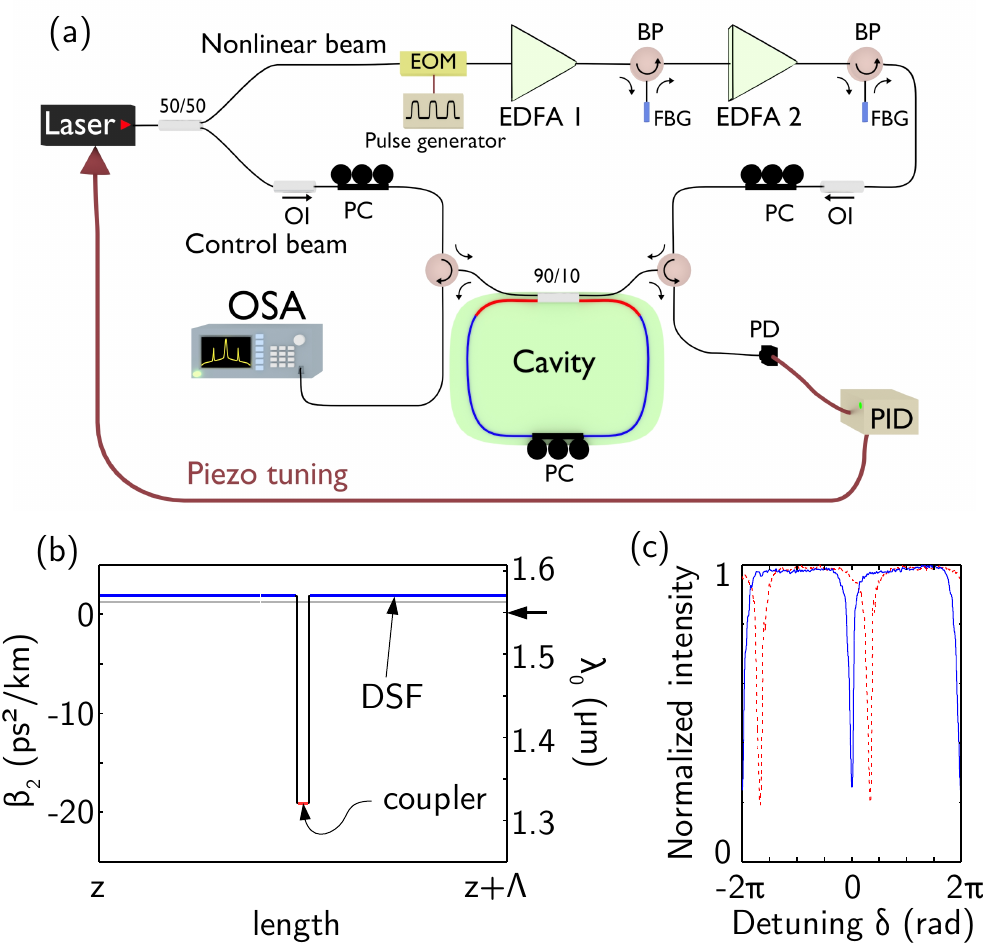}
\caption{\label{fig:disp} (a) Experimental setup \cite{sup_mat}. (b) GVD map of the cavity over one roundtrip, centered on the 90/10 SMF coupler. The gray horizontal line gives the average GVD $\beta_2^{av}$ at pump wavelength $1550~\text{nm}$ (arrow on the right vertical axis, calibrated in terms of wavelengths). (c) Normalized transfer functions of the cavity for the Control beam (dashed red) and the Nonlinear beam in the linear regime (blue).}
\end{figure}

\textit{Experiments}.-- We built a fiber ring cavity presenting the piecewise constant dispersion profile shown in Fig. \ref{fig:disp}(b). The ring has a total length of $51.6~\text{m}$, and is composed of a $50~\text{m}$ long, specially designed dispersion shifted fiber (DSF, with GVD $\beta_2 = 2~\text{ps}^2/\text{km}$) directly spliced to the two pigtails (total length $1.6~\text{m}$) of the input/output coupler made of a standard single-mode fiber (SMF-28 with GVD $\beta_2 = -19~\text{ps}^2/\text{km}$). The average nonlinear coefficient is $\gamma = 5.5~\text{/W/km}$. The cavity is pumped at $1550~\text{nm}$ (well below the average zero dispersion wavelength of $1562~\text{nm}$), where the values of GVD reported above gives a normal average dispersion $\beta_{2}^{av} \approx 1.35~\text{ps}^2/\text{km}$. The experimental setup is sketched in Fig. \ref{fig:disp}(a).

In order to validate the general behavior depicted in Fig. \ref{fig:mapping}(d), we contrast experiments made at relevant values of the normalized detuning, namely $\Delta = 4$ and $6.25$. 

\begin{figure*}
\includegraphics[width=\textwidth]{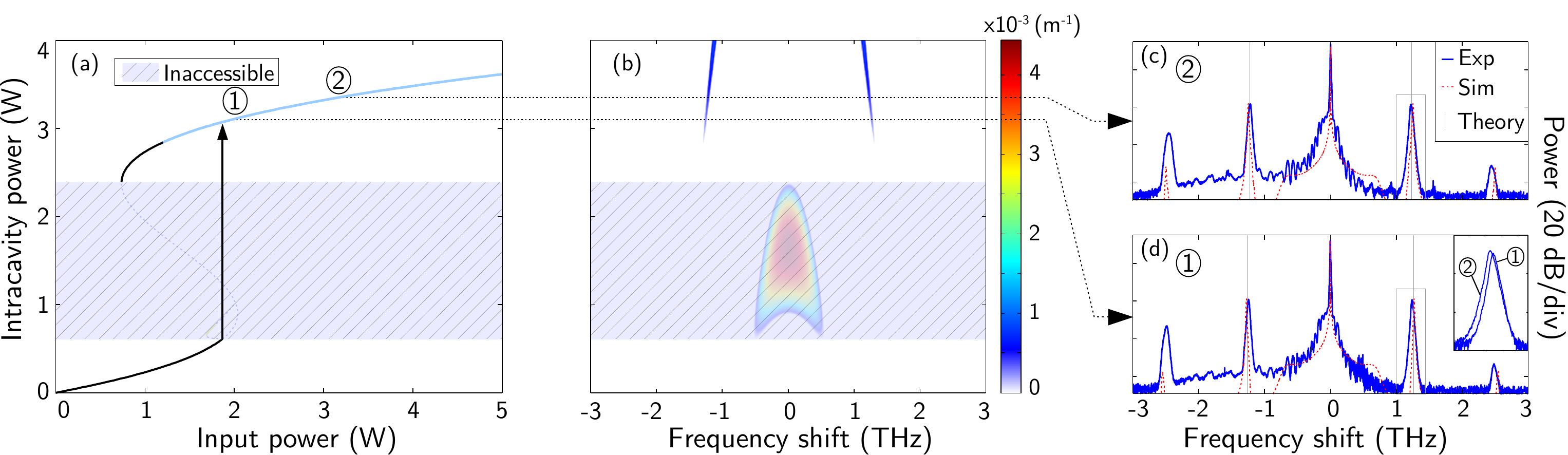}
\caption{\label{fig:exp1} (a) Bistable response of the cavity calculated for $\Delta = 4$ ($\delta = \pi/4.5~\text{rad}$, $\alpha = 0.175$); the hatched region is inaccessible. (b) Pseudo-color level plot of the gain spectrum as a function of the intracavity power calculated from Floquet  analysis \citep{conforti_modulational_2014}. (c-d) Comparison of experimental spectra (solid blue), spectra obtained from numerical integrations of the periodic LLE (\ref{eq:LLE}) (dashed red), and analytical estimates (vertical line, Theory) from Eq. (\ref{eq:QPM}) with $m=1$ for two different powers labelled 1 and 2 on the upper branch (see also Fig. \ref{fig:mapping}): (c) $P_{in} = 3.16~\text{W}$; (d) $P_{in} = 1.97~\text{W}$. Inset: close-up view of the sideband for the two powers. 
}
\end{figure*}

Figure \ref{fig:exp1} shows the results obtained for $\Delta = 4$ (vertical dashed line labelled A in Fig. \ref{fig:mapping}(d)).
For input powers below $1.7~\text{W}$ we do not observe any spectral signature of periodic structures in the output spectrum. Indeed the system is stable and simply follows the lower branch of the steady-state response shown in Fig. \ref{fig:exp1}(a). However, when the power exceeds the threshold for the unstable region (hatched region in Fig. \ref{fig:exp1}(a,b)), the system jumps on the upper branch. The latter is unstable owing to $m=1$ parametric (Faraday) resonance  as shown by the gain blue tongues in Fig. \ref{fig:exp1}(b). Consistently, we observe stable generation of Faraday (primary and harmonics) sideband pairs, as shown by the spectra reported in Fig. \ref{fig:exp1}(c,d), corresponding to two specific points (labelled 1 and 2) on the upper branch. These are in excellent agreement with spectra (dashed red curves) calculated from numerical simulations of the LLE (\ref{eq:LLE}). Moreover, the sidebands appear at $1.24$ and $1.22~\text{THz}$, for case 1 and 2, respectively, thus confirming the expected downshift for increasing power [see close-up inset in Fig. \ref{fig:exp1} (d)], in good agreement with the estimate from Eq. (\ref{eq:QPM}) which gives $1.26$ and $1.23~\text{THz}$, respectively (vertical grey lines). We also notice a strong amplitude asymmetry between the harmonics, which numerical simulations of the extended LLE allow us to attribute to the third-order dispersion that induces further symmetry breaking, as pointed out in uniform passive cavities \cite{leo_nonlinear_2013}. Finally notice that in this regime (i.e., $\Delta<4.25$), periodic Turing structures cannot be observed because they bifurcate subcritically, remaining unstable  in the relevant range of powers \cite{coen_bistable_1999}. 
Yet, the Turing instability still plays a crucial role, being responsible for inducing the upswitching at the input power $\sim 1.8$ W [obtained from the dimensionless threshold $P_u=\alpha$ arising from $g=0$ in Eq. (\ref{eq:Turing})], thus lowering the threshold for the formation of Faraday structures below the knee point $\text{P}_u^-$ [$\sim 2.1~\text{W}$, see Fig. \ref{fig:exp1}(a)].

\begin{figure*}
\includegraphics[width=\textwidth]{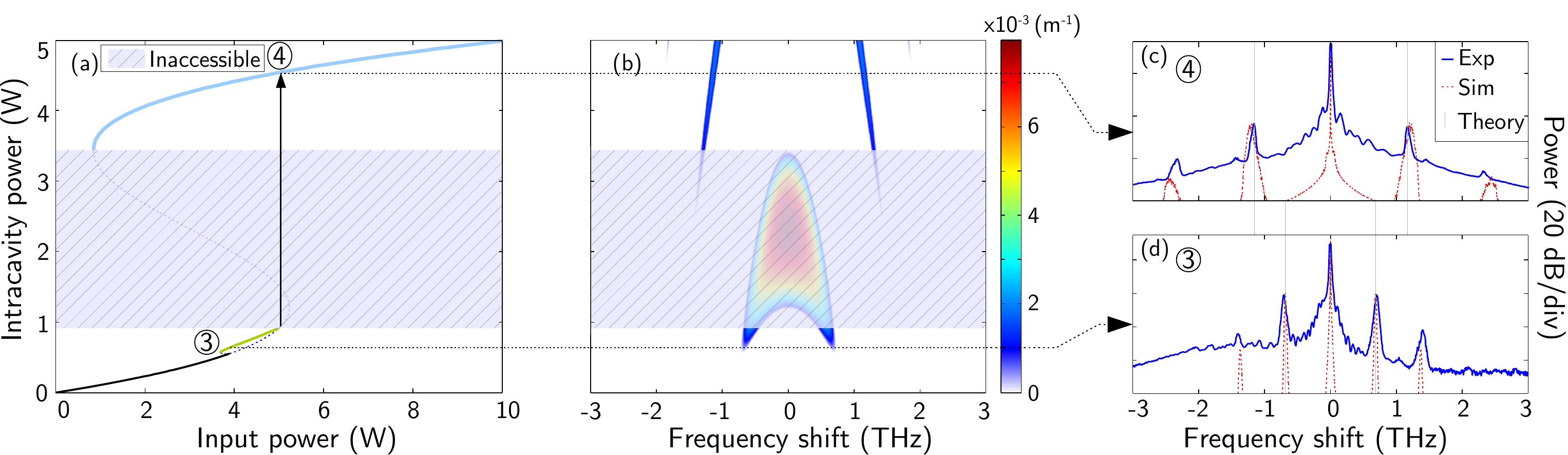}
\caption{\label{fig:exp2} As in Fig. \ref{fig:exp1} for $\Delta = 6.25$ ($\delta = \pi/3.2~\text{rad}$, $\alpha = 0.157$). Here the two spectra in (c-d) are relative to points (labelled 4 and 3) on two different (upper and lower) branches corresponding to: (c) $P_{in} = 5.02~\text{W}$ ($4.27~\text{W}$ in experiment); (d) $P_{in} = 3.9~\text{W}$ ($3.55~\text{W}$ in experiment). Estimated frequencies (Theory) are from Eq. (\ref{eq:QPM}) with $m=1$ in (c) and from Eq. (\ref{eq:Turing}) in (d).
}
\end{figure*}

A similar experiment is presented in Fig. \ref{fig:exp2} for $\Delta = 6.25$ (vertical dashed line labelled B in Fig. \ref{fig:mapping}(d)). At variance with previous case, the measured spectrum exhibits the stable formation of sidebands over the lower branch. An example is shown in Fig. \ref{fig:exp2}(d), where the primary sidebands are located at $0.70~\text{THz}$. This is consistent with the fact that, while the periodic solutions corresponding to Turing structures continue to bifurcate subcritically, a stable branch exist for a finite range of input powers, as shown by the green curve in Fig. \ref{fig:exp2}(a). Note that this range is quite limited despite the fact that the lower branch is significantly more extended in terms of input powers (compared with $\Delta=4$ case). Then, when the power exceeds the value where the Turing branch merge on the stationary response, the Turing instability induces up-switching towards the  upper branch. As described above, however, this branch presents narrowband Faraday instability [see Fig. \ref{fig:exp2}(b)] and hence two sidebands are still observed in the spectra [see Fig. \ref{fig:exp2}(b)], though at much larger frequency ($1.16~\text{THz}$). As can be seen, experimental spectra (blue curves) in Fig. \ref{fig:exp2}(c) and \ref{fig:exp2}(d) are in excellent agreement with numerical simulations (dashed red curves) and with the analytical predictions of the positions for the sidebands (vertical grey lines, $0.69~\text{THz}$ and $1.15~\text{THz}$ respectively). The large difference of frequency shifts between Fig. \ref{fig:exp2}(c) and \ref{fig:exp2}(d) allows us to claim that we have unambiguously observed the crossover between the two instabilities.

\begin{figure}
\includegraphics[width=.47\textwidth]{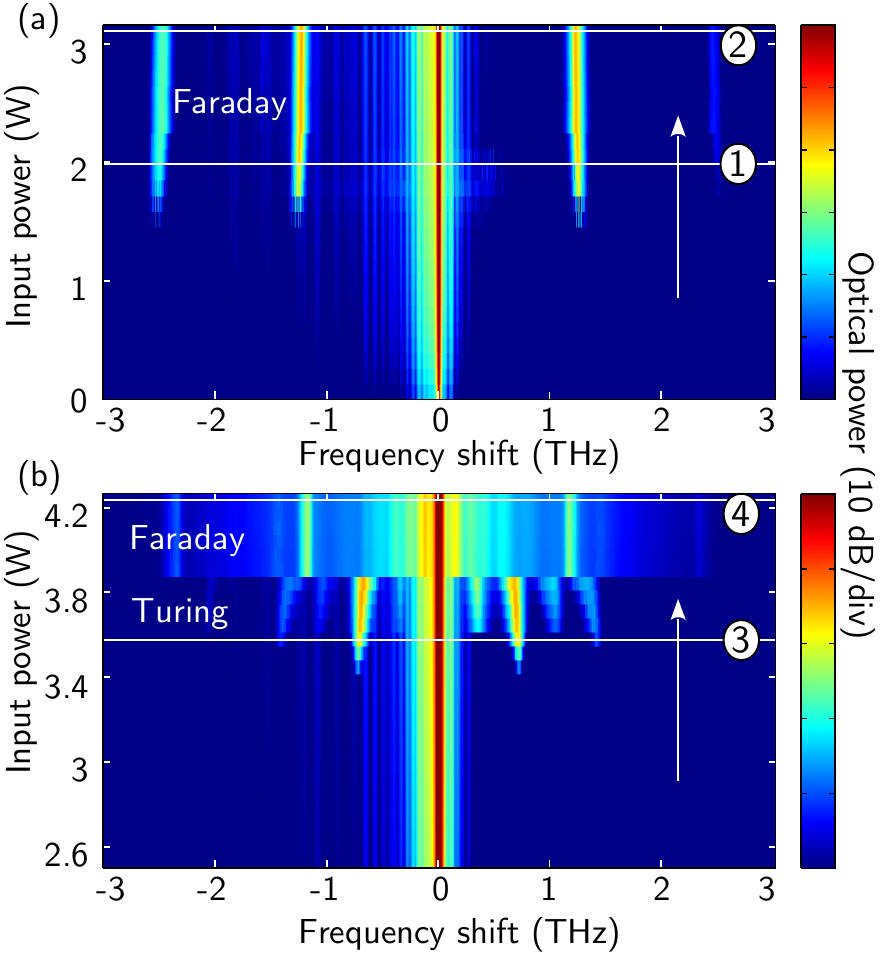}
\caption{\label{fig:sweepings} Experimental optical spectra at the cavity output vs. input power for (a) $\Delta = 4$ and (b) $\Delta = 6.25$. The numbered horizontal lines refer to the spectra shown in Figs. \ref{fig:exp1} and \ref{fig:exp2}.}
\end{figure}

Figure \ref{fig:sweepings} shows how the steady-state output spectrum changes when we adiabatically increase the input power. It clearly illustrates the two distinct scenarios of switching dynamics. In Fig. \ref{fig:sweepings}(a), for $\Delta = 4$, we observe the abrupt appearance of the Faraday instability sidebands when the input power exceeds $1.7~\text{W}$, which is in good agreement with the predicted switching threshold of $1.8~\text{W}$ [Fig. \ref{fig:exp1}(a)]. 
Conversely, for $\Delta = 6.25$, Fig. \ref{fig:sweepings}(b) first shows the power-induced tuning of the low-frequency Turing sidebands, until  eventually the abrupt switching to higher frequency sidebands which is the clear signature of the crossover to the Faraday branch. The Turing sidebands over the lower branch are observed in the range of input powers $3.4-3.9~\text{W}$, which is slightly lower than the theoretical expectation $3.9-5~\text{W}$ [see Fig. \ref{fig:exp2}(a)]. We attribute such larger discrepancy to the fact that the bistable response of the system is increasingly sensitive to environmental fluctuations because of the large detuning. Indeed, in this case, we consistently observe typically up to $15\%$ variations in the power threshold between repeated runs of the experiment.

It is important here to emphasize that all these spectra can remain stationary for about ten minutes which correspond to a few billion roundtrips. Until now, the evidence for the sidebands due to the Turing instability in a uniform passive fiber cavity were only given in the transient regime \cite{coen_competition_1999, coen_bistable_1999} or associated with  period-doubling dynamics \cite{coen_modulational_1997}. Our results thus constitutes the first experimental observation of stationary modulational instability spectra on both the lower and upper branches of the bistable response of a passive cavity.

\textit{Conclusions}.-- We have reported the first example of a bistable system which dynamics is dramatically affected by the excitation of modulated structures due to competing Turing and Faraday branches.
As our experiments unambiguously show, either the system can exhibit direct up-switching to Faraday 1D temporal patterns or crossover from Turing to Faraday modulated structures. 
These results demonstrate the feasibility of controlling the dynamics of a bistable system via periodic modulations. Besides being of interdisciplinary interest for non equilibrium systems, this could find immediate application for mode-locking, frequency comb and soliton generation in normally dispersive microresonators \cite{xue_mode-locked_2015, huang_mode-locked_2015, bao_stretched_2015}. 

\begin{acknowledgments}
This work was partly supported by IRCICA (USR 3380 Univ. Lille - CNRS), by the Agence Nationale de la Recherche (grants TOPWAVE, NoAWE, FOPAFE, Labex CEMPI and Equipex FLUX), by the French Ministry of Higher Education and Research, the Nord-Pas de Calais Regional Council and Fonds Europ\'{e}en de D\'{e}veloppement \'{E}conomique R\'{e}gional (grant CPER Photonics for Society). S.T. acknowledges also the grant PRIN 2012BFNWZ2. We are grateful to Laure Lago for providing the fiber Bragg gratings.
\end{acknowledgments}

\end{document}